# U.S. Data Center Growth Through 2030: A Critical Review of Grid Adequacy and Transmission Expansion Planning

Babak Porkar and Mona Ghassemi
Department of Electrical and Computer Engineering, The University of Texas at Dallas, Richardson, TX, USA
mona.ghassemi@utdallas.edu

**Abstract— The growth of large data centers raises questions about both supply adequacy and the suitability of load representations used in connection studies. This paper critically reviews the U.S. demand, generation, and transmission outlook through 2030 and examines the implications for transmission expansion planning (TEP). Comparison of public forecasts highlights distinctions among annual average demand, peak demand, and grid connection capacity. National generating-capacity totals cannot establish a dependable surplus because resource availability, regional adequacy, transmission constraints, and project timing determine supply capability. Recent grid events and published dynamic studies demonstrate that faults can trigger substantial data-center transfers to backup supply, followed by delayed or coordinated demand recovery. These behaviors require appropriate representation within the conventional planning framework, alongside assessment of the requested maximum grid import. The selected literature reveals limited integration of validated data-center dynamic models into the technical screening and economic comparison of expansion alternatives. The review identifies the need to establish whether fault response and recovery change feasible connection capacity, required reinforcement, or the preferred transmission investment.**

## I. INTRODUCTION

Data centers support banking, healthcare, cloud services, research, and artificial intelligence (AI). Their electrical demand includes computing, cooling, and power conversion. AI increases data-center electricity demand through model training, which develops AI models using data, and inference, which uses trained models to answer questions or make predictions. Conventional services, including cloud computing, banking, and video streaming, also require substantial electrical power [1, 2]. For power system planners, the relevant considerations are the data center facility's electricity demand and electrical behavior at its grid connection point, together with nearby projects and available network capacity. The growth of large data centers raises two related planning questions. Can available and planned generation and transmission serve their projected demand, and do conventional transmission expansion planning (TEP) studies adequately represent their response to grid disturbances? Answering the first question requires assessing regional resource availability and transmission deliverability, while answering the second requires load models that capture fault-induced demand transfers and subsequent reconnection. This paper critically reviews the current U.S. generation, transmission, and demand outlook through 2030, including projected data-center growth. It then examines research on data-center behavior during and after faults and identifies how these findings should be incorporated into conventional TEP studies. As a critical review, the paper evaluates the assumptions and limitations of existing studies and uses comparative analysis to highlight technical gaps in assessing data-center integration, particularly in load modeling, fault response, and the link between dynamic performance and transmission investment decisions.

## II. DEMAND AND GENERATION THROUGH 2030

### *A. Energy consumption and connection demand*

In its June 2026 report, Lawrence Berkeley National Laboratory (LBNL) estimates that U.S. data centers consumed 192 TWh of electricity in 2024, representing 4.7% of national electricity consumption and an annual average demand of approximately 21.9 GW. Approximate readings of the report's reference-case projection indicate 240 TWh in 2025 and 300 TWh in 2026, corresponding to annual average demands of approximately 27 GW and 34 GW, respectively. These values are graphically estimated rather than explicitly tabulated. The report treats 2024 as its last historical year because its equipment-shipment data extend only through the third quarter of 2025, so the 2025 results already include projections. For 2030, the reference case reaches 649 TWh, equivalent to approximately 74.1 GW of annual average demand and 11.8% of projected national electricity consumption. The combined uncertainty scenarios span 521–843 TWh, equivalent to approximately 59.5–96.2 GW and 9.5–15.3% of national consumption. These GW values represent annual averages, while connection planning requires maximum demand and its coincidence with other system loads [1]. Average power equals TWh × 1,000 divided by annual hours, using 8,784 h for 2024 and 8,760 h for 2025, 2026, and 2030. LBNL assumes that this average demand equals 50% of aggregate grid connection capacity. Under this assumption, its reference-case projection of 74.1 GW average demand implies approximately 148 GW of total data-center grid connection capacity in 2030, including existing and future facilities. This represents an estimated grid connection requirement, rather than installed IT equipment capacity. For TEP, each facility's requested maximum grid import must be assessed under applicable reliability criteria, accounting for connection location, commissioning schedule, and coincidence with other loads [1]. The Electric Power Research Institute (EPRI) estimates aggregate data-center peak demand of 21–22 GW in 2024 and projects 45, 71, and 94 GW under its low, medium, and high 2030 scenarios, respectively [3]. These are non-coincident totals, obtained by summing state-level peaks regardless of whether they occur simultaneously. EPRI's peak-demand forecasts therefore describe a different quantity from both LBNL's 74.1 GW annual average demand and its approximately 148 GW estimated aggregate grid connection capacity for 2030. EPRI includes cryptocurrency mining, whereas LBNL excludes it. Differences between the projections may also arise from assumptions about how heavily facilities operate, which planned projects become operational by 2030, and how quickly computing equipment is installed after a facility opens. For TEP, these national estimates provide context, while connection studies require each facility's requested maximum grid import and its location within the network [1], [3]. EPRI presents measured data-center load profiles showing sustained electricity demand at selected facilities [3]. For one single-tenant data center in EPRI's greater-than-75-MW capacity category, annual average and peak demand are approximately 75% and 80% of its reported nameplate capacity, respectively [3]. Its annual load factor—the ratio of average demand to peak demand—is approximately 94%, indicating that average demand is close to the observed peak. These facility-level observations differ from LBNL's assumed ratio of average demand to aggregate grid connection capacity and should not be treated as equivalent. The example is not identified as exclusively serving AI workloads and does not establish a universal planning factor. For TEP, assess the requested maximum grid import unless a lower limit is enforceable. Grid Strategies' November 2025 report provides a complementary perspective by aggregating utility forecasts of total U.S. electricity demand. It projects aggregate summer peak demand increasing from approximately 829 GW in 2025 to 995 GW in 2030, with data centers accounting for approximately 90 GW, or 55%, of the projected 166 GW increase [4]. These totals sum planning-area peaks and do not establish a simultaneous national peak. Unlike LBNL's annual average data-center demand [1] and EPRI's total data-center peak-demand scenarios [3], the 90 GW represents additional peak demand attributed to data centers between 2025 and 2030. Although conceptually closer to EPRI's peak-demand estimates, it requires a common baseline, coverage, and aggregation method for a meaningful comparison. However, market analyses reviewed by Grid Strategies suggest approximately 65 GW of additional data-center demand through 2030, compared with approximately 90 GW included in utility forecasts. This 25 GW difference suggests that utilities may overestimate how much data-center demand will materialize by 2030 [4]. These differences highlight the need to examine forecast definitions and project uncertainty before using national projections in planning studies.

### *B. Installed capacity and dependable supply*

The U.S. Energy Information Administration (EIA) reports 1,324.7 GW of utility-scale nameplate generating capacity and 1,230.4 GW of net summer capacity for 2024 [5]. Nameplate capacity is the sum of generators' rated capacities, whereas net summer capacity reflects their maximum output under summer conditions after subtracting electricity consumed within the plants. High ambient temperatures can reduce gas-turbine output, while warmer cooling water can constrain steam-cycle generation; pumps, fans, and other plant equipment also consume electricity. The resulting difference is 94.3 GW, approximately 7.1% of nameplate capacity, and should not be interpreted as capacity out of service. Net summer capacity nevertheless does not guarantee simultaneous availability during critical hours, because outages, weather-dependent generation, fuel availability, and other operating constraints must also be considered. EIA's *Annual Energy Outlook 2026 Counterfactual Baseline* projects approximately 1,452 GW of electric power sector net summer capacity in 2030 [6]. This total includes storage and excludes separately reported end-use generators; its coverage therefore differs from the 2024 utility-scale figures. The projected mix comprises approximately 782 GW of fossil, nuclear, hydrogen, and fuel-cell capacity; 89 GW of hydro and other renewables excluding wind and solar; 258 GW of solar; 213 GW of wind; and 111 GW of storage, including pumped storage [6]. Comparing the projected 1,452 GW of net summer capacity with approximately 995 GW of aggregate summer peak demand gives an apparent difference of 457 GW [4, 6]. However, this difference does not establish a dependable surplus or a valid planning reserve margin: the figures originate from separate forecasting frameworks, and net summer capacity does not represent the accredited contribution of all resources during critical hours. Dependable supply depends on weather, generator outages, fuel availability, hydro conditions, storage duration and recharge, and transmission deliverability. Annual capacity factors cannot establish resource availability during a specific critical hour. Adequacy assessment requires consistent regional demand and supply scenarios, chronological weather and storage modeling, correlated outages, transmission constraints, and explicit reliability criteria.

## III. Geographic and Transmission Constraints

### *A. Three major interconnections*

The contiguous United States has three major synchronous interconnections: Eastern, Western, and the Electric Reliability Council of Texas (ERCOT). ERCOT serves most of Texas, while other parts of the state belong to the Eastern or Western Interconnection. Alaska and Hawaii operate separate systems [7]. Although the three major interconnections operate at a nominal frequency of 60 Hz, they are not synchronized with one another. Existing links between them use high-voltage direct-current (HVDC) converters to control power transfers without requiring synchronization of the connected AC systems. Fig. 1 presents a separate regional assessment based on North American Electric Reliability Corporation (NERC) projections [8], rather than a disaggregation of the national forecasts in Section II. Its demand and resource totals differ because the sources use different forecasting assumptions, seasonal treatments, and resource-accounting conventions [4], [6], [8]. The resource totals include existing resources expected to remain available and planned additions classified by NERC as Tier 1 based on construction status, approved agreements, or qualifying resource plans; net imports from other assessment areas are excluded [8]. Eastern peaks are summer values, whereas the Western sum includes the Northwest's winter peak. The sums are non-coincident and follow regional conventions.

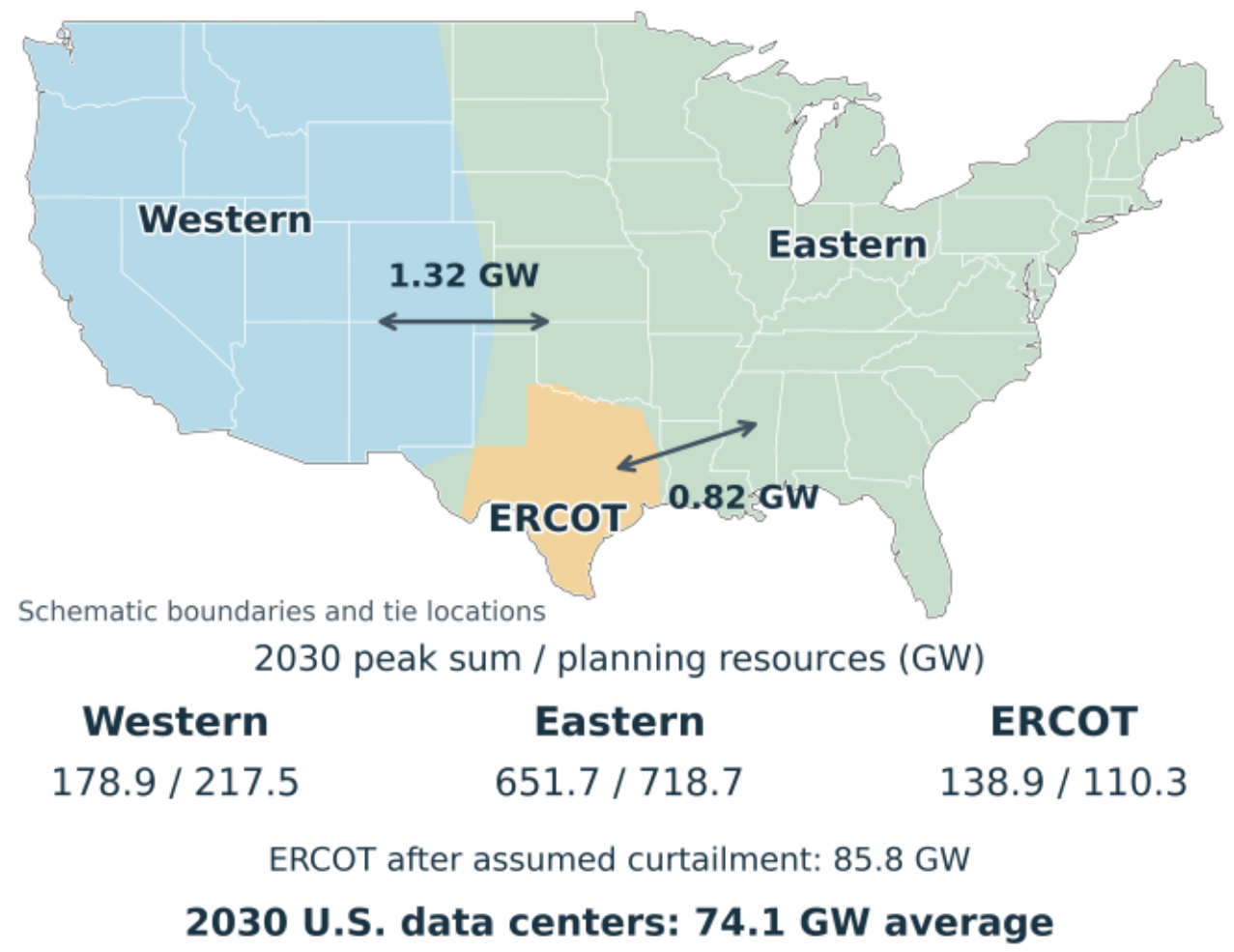

Fig. 1. Three major U.S. interconnections and 2030 demand–resource outlook [1], [7], [8], [11], [12]. Boundaries and locations are illustrative.

The totals in Fig. 1 imply apparent resource margins of approximately 10.3% for the Eastern Interconnection and 21.6% for the Western Interconnection. ERCOT instead shows a 20.6% deficit before demand reductions; its assumed 53.1 GW of demand response and curtailment reduces demand to 85.8 GW, producing an apparent margin of 28.6% [8]. These aggregate calculations do not establish resource adequacy: assessment-area peaks are non-coincident, the Western totals mix seasons, and transmission constraints can prevent resources in one area from supplying another. Moreover, positive margins must be evaluated against the applicable regional reliability benchmarks. For example, NERC projects 2030 reserve margins of 13.9% for PJM and 4.3% for MISO, both within the Eastern Interconnection and below their respective reference margins of 26.3% and 8.5% [8]. Thus, an apparent interconnection-wide surplus can coexist with adequacy concerns in individual assessment areas. A complementary geographic estimate comes from the BloombergNEF/DC Byte chart reproduced in the 2026 Sustainable Energy in America Factbook [9]. Approximate digitization gives 2025 facility demand of 23.4 GW in an Eastern grouping, 9.8 GW in a Western grouping, and 7.4 GW in ERCOT, or roughly 58%, 24%, and 18% of the U.S. total. These approximate market-region groupings do not exactly follow electrical boundaries. Their shares describe the distribution of data-center demand, not data centers' share of each grid's total consumption. These estimates should not be directly combined with the national projections because their definitions and geographic groupings differ.

### *B. Existing ties and planned projects*

Transmission expansion is also planned to support growing electricity demand. The U.S. Department of Energy (DOE) identifies a goal of increasing long-distance transmission capacity by approximately 16% by 2030, including construction of approximately 7,500 miles of new transmission lines [10]. Project completion, location, and network performance will determine how effectively these additions accommodate demand and generation growth. The seven domestic Eastern–Western back-to-back ties have an aggregate nominal rating of approximately 1.32 GW in the 2025 inventory [11]. They are Miles City, Rapid City, Stegall, Sidney, Lamar, Blackwater, and Eddy County. Their individual ratings are approximately 0.11–0.21 GW. The Eastern–ERCOT Oklaunion and Welsh–Monticello ties total 0.82 GW [12]. These are small relative to regional demand, and actual available transfers depend on outages and network conditions. Small installed ratings do not establish low utilization. Proposed projects could increase exchange, but their schedules matter. Three Corners Connector, an up-to-525 kV Eastern–Western HVDC project, identifies 2031 as its earliest operation [13]. North Plains Connector, also up to 525 kV, targets 2032 [14]. Southern Spirit proposes approximately 3 GW at ±525 kV between ERCOT and the Southeast, with a current target of 2032 [15]. Given their commissioning dates, these projects should be excluded from the 2030 baseline and considered beyond 2030. Within each interconnection, corridor limits and substation capability remain decisive even if boundary ties expand.

## IV. What Changes in Data Center Connection Planning

### *A. The established study framework*

From a transmission expansion planning perspective, a large data center represents a concentrated electrical load requiring an appropriate connection and sufficient network capability to supply its requested maximum demand. The facilities considered here have demands of tens to hundreds of megawatts, while large campuses may reach gigawatt scale at full development. Such loads often require transmission-level supply through dedicated substations. The connection voltage and reinforcement requirements depend on the facility's demand, local network conditions, and applicable reliability criteria.

A first planning question is which demand should determine the required connection capacity: nameplate capacity, peak demand, or annual average demand. The relevant basis is the maximum grid import requested by the customer, rather than equipment nameplate capacity or average consumption alone. EPRI's measured example exhibits peak demand of approximately 80% of reported nameplate capacity [3], illustrating a potentially consequential distinction for large facilities. For an 800 MW facility, an agreed 640 MW maximum grid import would reduce the connection requirement by 160 MW and could materially change feasible transmission reinforcement alternatives. However, an observed utilization level does not establish an enduring demand limit. Planning for the lower import requires an enforceable arrangement supported by adequately rated on-site resources or load curtailment, with their availability, operating duration, and control performance assessed. Otherwise, the connection must accommodate the requested maximum demand, including infrequent high-demand periods. Conventional large-load connection planning evaluates alternative supply schemes using a future-year network model representing the study area and its interaction with the wider system. Although proposed reinforcements often concentrate near the connection point, the assessment addresses both local and wider network impacts. Each alternative undergoes load-flow analysis under the normal operating condition and applicable contingencies, including single-element outages and other events required by the relevant planning criteria [16]. Studies consider system peak-load and relevant off-peak conditions. An illustrative light-load case may set surrounding demand at 40% of its peak value while maintaining the facility at its maximum demand, testing whether the network can accommodate full facility operation during periods of low system demand. Alternatives satisfying steady-state thermal and voltage requirements undergo short-circuit and transient-stability assessments. Short-circuit calculations, including three-phase and single-line-to-ground faults, determine whether fault duties remain within circuit-breaker interrupting capabilities and other equipment ratings. Transient-stability studies evaluate whether generators remain in synchronism following specified disturbances, including three-phase faults with appropriate clearing sequences, and whether the system meets applicable dynamic performance criteria [16]. Technically acceptable alternatives are then compared based on investment costs and total network power losses. The same planning framework applies to data-center connections. However, recent grid events highlight the need for targeted assessment of data-center fault response and demand recovery within this framework. The following subsection examines these behaviors and their implications for connection planning.

*B. Fault response and demand recovery*

Recent grid events demonstrate that a fault in the power system can cause data centers to transfer to backup supply, substantially reducing their electricity demand from the grid while maintaining facility operation. On July 10, 2024, a disturbance involving a 230 kV transmission line in the Eastern Interconnection caused approximately 1,500 MW of data-center demand to transfer away from the grid [17]. PJM subsequently reported another approximately 1,500 MW transfer in the Dominion zone on February 17, 2025 [18]. A larger event occurred on July 22, 2026, when a normally cleared 230 kV line fault in Northern Virginia triggered successive data-center transfers to backup supply, producing an overall grid-demand reduction of approximately 3,800 MW. The resulting generation–demand imbalance caused voltage and frequency excursions. Operators reduced generation and adjusted reactive-power resources to restore operating conditions; PJM reported no significant reliability impacts [18]. These events demonstrate that clearing the initiating fault does not necessarily restore pre-fault data-center demand from the grid, highlighting the need to represent data-center protection, transfers to backup supply, and subsequent demand recovery in connection studies. Post-contingency load-flow analysis evaluates the network after the faulted element has been disconnected, but does not capture the voltage dip during the fault or the resulting data-center protection and control actions. At 60 Hz, an illustrative fault-clearing time of two to five cycles corresponds to approximately 33–83 ms. Although brief, this disturbance may cause a data center to transfer some or all of its demand to backup supply. Acceptable post-contingency bus voltages therefore do not establish that data-center demand will remain connected during the fault or immediately return after clearing. This behavior depends on disturbance magnitude and duration, converter controls, protection settings, and reconnection logic [19, 20]. A dynamic model and converter-based emulator are presented in [19]. In one tested case, a voltage reduction exceeding 30% triggered a transfer to battery operation after approximately 20 ms, maintaining power to the servers while reducing demand from the grid. This result illustrates how data-center demand can change within the fault-clearing interval, although the transfer threshold and delay reflect the tested configuration rather than universal data-center settings. A Pacific Northwest National Laboratory (PNNL) report further identifies converter current limits, synchronization controls, and protection settings as factors influencing fault ride-through [20]. Planning models should therefore represent the relevant computing and cooling loads, uninterruptible power supply (UPS) systems, and backup-supply transfer logic. When substantial data-center demand remains disconnected from the grid after fault clearing, generation may temporarily exceed the remaining demand and network losses. This imbalance can accelerate synchronous generators and raise system frequency, while changes in active and reactive demand can also produce local overvoltages. The subsequent return of data-center demand introduces another disturbance, whose severity depends on the amount reconnecting, the delay, and the rate of demand recovery. Similar protection settings across nearby facilities may cause correlated demand reductions or reconnections, increasing the aggregate impact. These effects are not exclusive to data centers, but their scale and concentration make them particularly relevant to connection planning. Dynamic studies should therefore assess the initiating fault, the resulting data-center demand transfer, and the subsequent recovery as a connected sequence [17, 18, 20]. NERC's May 2026 draft guideline similarly emphasizes site-specific modeling, interconnection performance, and model validation [21]. It is cited here as draft guidance, rather than an enforceable ride-through standard.

*C. Research Findings and Technical Gaps*

The reviewed literature addresses two largely separate questions: how additional data-center demand affects infrastructure investment, and how data-center electrical systems respond to disturbances. The selected expansion studies primarily represent data centers as prescribed demands, whereas dynamic studies examine protection, backup-supply transfers, and reconnection without selecting transmission investments. The principal gap within these studies is the limited demonstrated integration of data-center dynamic behavior into the technical screening and economic comparison of connection alternatives. A co-optimized generation and transmission expansion study considers an additional 500 MW of fixed data-center demand on a modified IEEE RTS-24 network [22]. Another study optimizes hybrid AC/DC transmission additions with a 50 MW data center on a modified Garver system [23]. Both address infrastructure investment, but neither demonstrates how fault-induced demand transfers and subsequent reconnection affect the selected plans. Their results illustrate the consequences of concentrated demand under specified assumptions; they do not establish that the selected alternatives satisfy the full connection-study process described in Section IV.A. A dynamic model comprising UPS systems, cooling motors, and computing loads is applied to a 1,479-bus representation of the all-island Irish transmission system in [24]. The initial facility includes 300 MW of server capacity and 60 MW of cooling demand. In a larger, approximately 420 MW case, rapid demand restoration produces repeated disconnection and reconnection, termed *flapping*. Staggering the return of ten UPS blocks mitigates this behavior, while soft starting further smooths demand recovery. These results demonstrate that connection acceptability can depend on the restoration sequence as well as the initial demand transfer. However, the findings depend on assumed protection and control settings; applying them to U.S. systems requires validation of facility parameters and system response. An electrical and thermal model of a 10 MW data center is evaluated on a WSCC 9-bus system in [25]. Simulations consider faults lasting 83, 150, and 250 ms, together with a separate voltage ride-through case involving 0.45 p.u. voltage for 0.3 s. The results illustrate that a brief voltage depression does not necessarily cause immediate disconnection. Together with the battery-transfer behavior reported in [19], they demonstrate why a universal data-center trip threshold is inappropriate. Protection settings, converter controls, and facility composition determine the response. Nevertheless, component validation and small-system demonstrations do not establish the aggregate behavior of a campus with hundreds of megawatts of demand. Centralized and decentralized controls for improving data-center low-voltage ride-through are examined through integrated transmission-system and internal facility simulations in [26]. This preprint indicates that internal control changes can improve disturbance response, providing a potential mitigation option alongside network reinforcement. It does not, however, determine a preferred transmission expansion plan or quantify investment savings under equivalent service and reliability requirements. These findings suggest that TEP studies should test whether validated data-center models change the technical acceptability or economic ranking of connection alternatives. Relevant assessments should cover peak and light-load conditions, applicable fault-clearing and reclosing sequences, and correlated demand transfers and recovery across nearby facilities [16], [20], [21]. Positive-sequence simulations can assess system-level electromechanical response, supplemented by electromagnetic transient studies where converter controls, phase jumps, or unbalanced disturbances require greater detail [20]. Where unacceptable performance occurs, planners should evaluate network reinforcement and permissible changes to facility protection or controls under consistent import requirements and reliability obligations. Research should quantify whether these behaviors change feasible connection capacity, required reinforcement, or preferred investment within conventional planning.

## V. Conclusions

This critical review examines the U.S. demand, generation, and transmission outlook through 2030 and its implications for integrating large data centers. Public projections indicate substantial growth, but annual average demand, peak demand, and grid connection capacity describe different quantities and cannot be used interchangeably. Likewise, national generating-capacity totals do not establish a dependable surplus. Resource availability, regional adequacy, transmission constraints, and project commissioning schedules determine whether supply can reach new loads. Data-center connections remain subject to the conventional TEP framework, with the requested maximum grid import providing the demand basis unless a lower limit is enforceable. Recent events and dynamic studies nevertheless demonstrate the importance of representing fault-induced transfers to backup supply and subsequent demand recovery. Acceptable post-contingency voltages alone cannot establish acceptable performance throughout this sequence. The principal technical gap identified in the selected literature is the limited integration of validated data-center dynamic models into the screening and economic comparison of expansion alternatives. Addressing this gap requires realistic future-year network studies that establish whether these behaviors change feasible connection capacity, required reinforcement, or the preferred investment.

## References

[1] S. J. Smith et al., "United States Data Center Energy Usage Report: 2025 Update," Lawrence Berkeley National Laboratory, LBNL-2001758, Jun. 2026, doi: 10.71468/P1RP4F.

[2] International Energy Agency, "Energy and AI," Paris, France, 2025. [Online]. Available: https://www.iea.org/reports/energy-and-ai

[3] Electric Power Research Institute, Powering Intelligence 2026, Feb. 2026, Fig. 3 and peak-load projections. [Online]. Available: https://powering-intelligence.epri.com/annual-peak-use.html

[4] J. D. Wilson, S. Meyer, Z. Zimmerman, and R. Gramlich, Power Demand Forecasts Revised Up for Third Year Running, Led by Data Centers, Grid Strategies, Nov. 2025. [Online]. Available: https://gridstrategiesllc.com/wp-content/uploads/Grid-Strategies-National-Load-Growth-Report-2025.pdf

[5] U.S. EIA, Electric Power Annual 2024, Table 4.3. [Online]. Available: https://www.eia.gov/electricity/annual/archive/2024/table.php?t=epa_04_03.html

[6] U.S. EIA, Annual Energy Outlook 2026, Counterfactual Baseline, Tables 9 and 16, Apr. 2026. [Online]. Available: https://www.eia.gov/outlooks/aeo/

[7] U.S. EIA, "U.S. electric system is made up of interconnections and balancing authorities," Jul. 2016. [Online]. Available: https://www.eia.gov/todayinenergy/detail.php?id=27152

[8] NERC, 2025 Long-Term Reliability Assessment, Jan. 2026, regional dashboards and reserve margins. [Online]. Available: https://www.nerc.com/globalassets/our-work/assessments/nerc_ltra_2025.pdf

[9] BloombergNEF and BCSE, 2026 Sustainable Energy in America Factbook, p. 59, Feb. 2026. [Online]. Available: https://bcse.org/wp-content/uploads/2026/02/2026-Sustainable-Energy-in-America-Factbook.pdf

[10] U.S. Department of Energy, "Grid deployment and transmission." Accessed: Nov. 9, 2026. [Online]. Available: https://www.energy.gov/topics/grid-deployment-and-transmission

[11] Southwest Power Pool, "West/East DC ties," Legal Forum presentations, Jun. 3, 2025. [Online]. Available: https://www.spp.org/documents/74016/june%203%202025%20spp%20legal%20forum%20presentations.pdf

[12] ERCOT, Report on Existing and Potential Electric System Constraints and Needs, Dec. 2023, HVDC transmission section. [Online]. Available: https://www.ercot.com/gridinfo/planning

[13] Grid United, "Three Corners Connector," project description and status. [Online]. Available: https://threecornersconnector.com/

[14] Grid United and ALLETE, "North Plains Connector," project description and status. [Online]. Available: https://northplainsconnector.com/

[15] Pattern Energy, "Southern Spirit Transmission," project description and schedule. [Online]. Available: https://patternenergy.com/projects/southern-spirit-transmission/

[16] NERC, TPL-001-5.1 Transmission System Planning Performance Requirements, requirements R1–R4 and Table 1. [Online]. Available: https://www.nerc.com/globalassets/standards/reliability-standards/tpl/tpl-001-5.1.pdf

[17] NERC, Incident Review—Considering Simultaneous Voltage-Sensitive Load Reductions, Jan. 8, 2025, "Load Details," Figs. 8–9. [Online]. Available: https://www.nerc.com/globalassets/our-work/reports/event-reports/incident_review_large_load_loss.pdf

[18] PJM, "PJM, Dominion review large load transfer event," Aug. 11, 2026. [Online]. Available: https://insidelines.pjm.com/pjm-dominion-review-large-load-transfer-event/

[19] J. Sun, S. Wang, J. Wang, and L. M. Tolbert, "Dynamic model and converter-based emulator of a data center power distribution system," *IEEE Trans. Power Electron.*, vol. 37, no. 7, pp. 8420–8432, Jul. 2022, doi: 10.1109/TPEL.2022.3146354.

[20] B. A. Ross and J. Follum, Electromagnetic Transient Modeling of Large Data Centers for Grid-Level Studies, PNNL-38817, Alpha Release, Dec. 2025, pp. 33–34. [Online]. Available: https://www.energy.gov/sites/default/files/2026-01/Data_Center_EMT_Models.pdf

[21] NERC, Draft Reliability Guideline—Risk Mitigation for Emerging Large Loads, May 2026. [Online]. Available: https://www.nerc.com/globalassets/who-we-are/standing-committees/rstc/reliabilityguideline_riskmitigationforemerginglargeloads.pdf

[22] H. Wan, X. Li, and J. Kwon, "Impact of data center load on co-optimized generation and transmission expansion planning with embedded unit commitment," *arXiv:2610.18385*, Oct. 2026. [Online]. Available: https://arxiv.org/abs/2610.18385

[23] M. Ahanch, M. Sanjabiasasi, R. McCann, and S. Canas Rivera, "Considering bulk data centers in hybrid AC/DC transmission network expansion planning," in *Proc. INTCEC*, 2025, doi: 10.1109/INTCEC65580.2025.11255733.

[24] A. Jiménez-Ruiz and F. Milano, "Data centre model for transient stability analysis of power systems," *IET Gener. Transm. Distrib.*, vol. 20, art. e70381, 2026, doi: 10.1049/gtd2.70381.

[25] P. P. Gyang, P. Chakraborty, L. Meegahapola, and X. Yu, "Dynamic modeling of a data center for power system stability studies," *IEEE Trans. Power Syst.*, vol. 41, no. 3, pp. 2317–2332, 2026, doi: 10.1109/TPWRS.2025.3639163.

[26] Y. Xie, W. Cui, and A. Wierman, "Enhancing data center low-voltage ride-through," *arXiv:2510.03867*, Oct. 2025, doi: 10.48550/arXiv.2510.03867.